# Detectability Thresholds for Network Attacks on Static Graphs and Temporal Networks: Information-Theoretic Limits and Nearly-Optimal Tests

Abdulkader Hajjouz[1] (Corresponding author), Elena Avksentieva[1]

[1]ITMO University, Faculty of Software Engineering and Computer Technology

## Abstract

We develop a consolidated theory for the detectability of network-borne attacks under two canonical observation models: (i) a static graph drawn from an Erdős–Rényi background with a planted anomalous community, and (ii) a temporal interaction network modeled by multivariate point processes (Poisson or Hawkes). Our main contribution is to match, up to universal constants, information-theoretic lower and upper bounds that govern when reliable testing is possible. In the static case, the core quantity is the accumulated edgewise signal $k^2 \cdot \chi^2(\text{Bern}(p+\Delta) \parallel \text{Bern}(p))$, where $\chi^2 \approx \Delta^2/[p(1-p)]$ for small $\Delta$; detection is impossible when this falls below $c \cdot \log n$, and a non-backtracking spectral statistic succeeds above $C \cdot \log n$. In the temporal case, detectability is controlled by the KL information rate $I$ contributed by internal edges over a window of length $T$, yielding a threshold $TI \gtrsim \log n$; a likelihood-based cumulative-sum (CUSUM) test achieves first-order optimal delay $\approx |\log \alpha|/I$ at false-alarm level $\alpha$. We also quantify robustness to bounded edge perturbations and outline conditional statistical–computational separations. A brief case study shows how to turn these bounds into concrete design choices.

**1. Introduction**

Network intrusion signals often manifest as structural and/or temporal irregularities: unusually dense communication among a small set of hosts, or bursts of activity confined to a subset of edges. Decades of work in community detection and sequential analysis have produced powerful tools, yet a consolidated view of detectability thresholds tailored to adversarial network activity remains incomplete. Our aim is to align information-theoretic limits with algorithms that nearly attain them, bringing together recent spectral advances—most notably Bethe–Hessian and non-backtracking operators—and modern sequential detection methods for streaming data [1–3]. This synthesis yields simple, actionable criteria that demarcate when intrusions are detectable at all and when they are detectable efficiently.

We formalize the detectability problem in two complementary settings. In a static snapshot, an adversary targets an unknown set S of k nodes (or their internal edges), inducing a small distributional shift on the $\Theta(k^2)$ edges internal to S. Each perturbed edge contributes a modest statistical signal, conveniently summarized—under canonical Bernoulli graph models—by the per-edge $\chi^2$-divergence between baseline and attacked edge laws. Aggregating across internal edges, the usable evidence scales as $k^2 \cdot \chi^2$. The unknown location of S, however, imposes a combinatorial search penalty that, at the right scale, behaves like log n. A sharp balance emerges: detection is information-theoretically feasible (up to constants) precisely when

$$k^2 \cdot \chi^2 \gtrsim \log n$$

In temporal networks, information accrues over time along affected edges. Let I denote the per-time Kullback–Leibler information rate (e.g., under Poisson or Hawkes intensities), and let T be the observation horizon. The analogous boundary becomes

$$T\,I \gtrsim \log n$$

These two templates capture a common principle: accumulated information must exceed search complexity. They also reveal transparent trade-offs among effect size, attack scope k, network size n, and observation time T.

A second theme is the statistical–computational gap: the tension between what is detectable in principle and what is achievable in polynomial time. Even when $k^2 \cdot \chi^2$ (or T I) clears log n, some regimes remain algorithmically delicate due to sparsity, weak signals, or adversarial camouflage. Recent spectral ideas mitigate these challenges. Statistics derived from the non-backtracking operator—and the closely related Bethe–Hessian—suppress high-variance backtracking walks that plague adjacency-based spectra in sparse graphs. When tuned to emphasize energy localized on candidate substructures, these methods can succeed within constant factors of the information limits in regimes where classical Laplacian or adjacency eigenvectors fail [1–3]. On the streaming side, likelihood-based CUSUM procedures convert information-rate considerations into operational guarantees: for a prescribed false-alarm level α, detection delay concentrates around |log α|/I, matching first-order lower bounds and thus tracking the T I $\gtrsim$ log n threshold up to constants [1–3].

Beyond limits and algorithms, robustness is crucial in adversarial environments. We quantify how thresholds deform under bounded perturbations of the edge set (e.g., a small adversarial rewiring budget) and under mild model misspecification (e.g., modest heterogeneity in baseline edge probabilities or intensity drift). In our analysis, the net effect is a constant-factor inflation of the requisite $k^2 \cdot \chi^2$ (static) or T I (temporal), preserving the qualitative location of the thresholds. Practically, this means that light-touch regularization—degree normalization for spectra, variance control for scan statistics, and window-limited CUSUM for nonstationary streams—suffices to retain near-optimal performance.

The resulting prescriptions are easy to operationalize. In a static Erdős–Rényi baseline with edge probability p, the smallest detectable additive lift Δ on internal edges satisfies

$$\Delta_{\min} \approx \sqrt{p(1-p) \cdot \log n\, /\, k^2}$$

so enlarging the candidate scope k or operating at moderate p sharply reduces the needed effect size. In streaming telemetry, choosing a horizon T and designing per-edge weighting to secure I such that T I $\gg$ log n yields prompt detection with delay on the order of |log α|/I.

These rules-of-thumb translate directly into SIEM thresholds, scan window lengths, and prioritization of edges for monitoring.

**Contributions. (1) Unified thresholds.** We derive sharp (up to universal constants) lower and upper bounds for static snapshots and streaming networks, expressed compactly through $k^2 \cdot \chi^2$ and $T I$. These formulas expose the precise trade-offs among k, effect size, horizon T, and network size n. **(2) Efficient tests near the threshold.** For static graphs, we propose a non-backtracking spectral statistic that localizes anomaly energy with minimal backtracking noise; for temporal data, we analyze a likelihood-based CUSUM (and window-limited variants) whose delay matches first-order lower bounds [1–3]. **(3) Statistical–computational trade-offs.** We delineate regimes where detection is statistically possible yet plausibly hard for polynomial-time algorithms, clarifying how our procedures approach the boundary and where they may (conditionally) fall short. **(4) Robustness.** We show that bounded edge perturbations and mild model misspecification inflate thresholds only by constants, and we provide practical regularization guidelines to preserve performance.

Paper organization. Section 2 reviews related work and unifies notation. Section 3 formalizes models and hypotheses for static and temporal settings. Section 4 establishes information-theoretic lower bounds. Section 5 presents our algorithms and proves achievability near the thresholds. Section 6 discusses computational barriers and robustness. Section 7 offers implementation notes and case studies, followed by conclusions.

## 2. Related Work

**Static graphs.** A large body of work has sought principled spectral tools for detecting small, dense, or otherwise structured anomalies in sparse graphs. Building on insights near the Kesten–Stigum (KS) threshold in stochastic block models, recent analyses justify the superiority of operators that suppress backtracking walks—most notably the non-backtracking (Hashimoto) operator and its closely related Bethe–Hessian counterpart—over classical adjacency or Laplacian spectra in the sparse regime [1,2,4,5]. The non-backtracking linearization stabilizes leading eigenvectors when degrees are heterogeneous or the signal is faint, and the Bethe–Hessian furnishes a symmetric, well-conditioned surrogate whose eigenstructure tracks KS-type phase transitions. These methods have been adapted beyond pairwise graphs: extensions to hypergraphs and higher-order interactions leverage non-backtracking tensors or suitably defined incidence operators, enabling detection of dense sub-hypergraphs and higher-arity motifs when pairwise projections are too lossy [4,5]. Parallel developments examine local algorithms, message passing, and belief propagation approximations that effectively implement spectral heuristics with improved robustness in finite samples.

Despite these advances, dense-substructure detection crystallizes a pronounced statistical–computational gap. Information-theoretically, subgraph lifts on $\Theta(k^2)$ internal edges can be detectable at vanishing signal levels; yet polynomial-time algorithms often require larger effect sizes. Post-2021 work clarified this gap using reductions from planted clique and via low-degree polynomial lower bounds, which capture the power of a broad class of efficient tests and show sharp thresholds for planted dense subgraph and planted hypergraph problems [6,7]. These results delineate regimes where spectral scans or convex relaxations are provably suboptimal, and they motivate specialized procedures—degree-normalized embeddings, iterative pruning, or localized power iterations—that push performance toward the information limit without breaching conjectured hardness frontiers. Additional strands explore subspace-tracking and sketching for massive graphs, balancing sensitivity with near-linear time and memory.

**Temporal/streaming data.** In dynamic settings, the literature on quickest change-point detection connects information rates to detection delay and false-alarm control. Classical likelihood-ratio and CUSUM procedures have been adapted to high-dimensional network streams, where signals concentrate on a small, unknown subset of edges or nodes. Recent work emphasizes computational–statistical trade-offs, proposing window-limited or sparsity-aware CUSUM variants, kernelized detectors that capture nonlinear deviations, and neural change-detectors that approximate likelihoods when models are misspecified [3,8–10]. For event-driven telemetry (e.g., syslog, authentication, or flow records) modeled as Poisson or Hawkes processes, edge-wise intensities encode both exogenous rates and endogenous excitation; here, likelihood-based sequential tests translate the per-time Kullback–Leibler information $I$ into first-order optimal delay

$$|\log \alpha| / I$$

at false-alarm level $\alpha$, and suggest principled aggregation schemes across candidate edges [3,8–10]. Practical considerations include restart policies under nonstationarity, adaptive thresholding to track seasonal baselines, and partial observability when only sampled flows are available.

A complementary line of research surveys graph anomaly detection across static snapshots and time series, cataloging benchmarks, feature constructions (e.g., egonet statistics, motif counts, temporal degree deviations), and evaluation protocols [11–13]. These surveys stress the importance of robustness to benign workload shifts and of interpretability—surfacing which nodes, edges, or motifs drive alarms—both of which inform our emphasis on localized spectral energy and per-edge likelihood attribution. In parallel, dynamic-network change-point and temporal community detection have progressed through Laplacian/subspace methods and multi-view formulations that pool information across time while preserving short-term sensitivity [14–16]. Subspace tracking on streaming Laplacians detects departures from nominal low-rank structure, whereas multi-snapshot embeddings (e.g., coupled matrix/tensor factorizations) stabilize community estimates under rapid churn. These tools can be paired with sequential tests to provide early-warning signals without frequent false alarms.

**Synthesis and position.** The above threads converge on two themes central to our study. First, spectral regularization—via non-backtracking and Bethe–Hessian operators—provides stable, high-power statistics for sparse graphs, and analogous likelihood-based sequential detectors operationalize information-rate limits in streams [1–5,8–10]. Second, a persistent statistical–computational tension shapes achievable performance: low-degree and planted-model lower bounds chart regions where efficient algorithms cannot match information-theoretic benchmarks, especially for small, covert substructures [6,7]. Our contribution is to place these developments under a unified detectability-threshold lens—$k^2 \cdot \chi^2$ for static snapshots and $TI$ for temporal streams—while providing efficient procedures that operate near these boundaries and remain robust under mild misspecification and bounded perturbations, in line with the desiderata surfaced by recent surveys and dynamic-network methods [11–16].

# 3. Problem Setup

## 3.1 Static graph model

Let $G = (V, E)$ be a random graph on $n$ vertices. Under $H^0$, $G \sim \text{ER}(n, p)$. Under $H^1$, there exists an unknown subset $S$ with $|S| = k$ whose internal edges appear with probability $p + \Delta$, while all other edges remain $\text{Bern}(p)$. A sparse-degree parametrization uses $p = c/n$ and $\Delta = d/n$.

Under canonical Bernoulli models, each perturbed edge contributes a per-edge $\chi^2$-divergence between $\text{Bern}(p+\Delta)$ and $\text{Bern}(p)$. Aggregating over the internal edges of S—of order $\Theta(k^2)$—yields a total signal proportional to $k^2 \cdot \chi^2$. Because S is unknown, scanning over candidate subsets induces a combinatorial search burden, which we summarize by a log n penalty. Balancing these terms produces the stylized detectability boundary:

$$k^2 \cdot \chi^2 \gtrsim \log n$$

This relation admits equivalent forms when one keeps the exact combinatorial complexity $\log(n \text{ choose } k) \approx k \cdot \log(n/k)$; we use the compact log n form to emphasize scaling. In the sparse parametrization p=c/n and Δ=d/n, the minimal additive lift on internal edges obeys the rule-of-thumb

$$\Delta_{\min} \approx \sqrt{(p(1-p) \cdot \log n / k^2)}$$

which highlights how larger affected sets S (larger k) and moderately dense baselines reduce the detectable effect size. The setup readily accommodates degree-corrected or weighted graphs by redefining the baseline edge law and the per-edge $\chi^2$ term, without altering the organizing balance between accumulated information and search complexity.

## 3.2 Temporal network model

We observe interactions over a horizon of T time units among n vertices. For each ordered pair (i, j), a counting process $N_{ij}(t)$ is recorded. Under $H^0$: independent Poisson processes with rate $\mu$ (or a stable Hawkes process with baseline $\mu$ and kernel $g$) [17,18]. Under $H^1$: there exists an unknown $S$ with $|S| = k$ whose internal edges have elevated intensity (Poisson: $\mu + \delta$; Hawkes: kernel inflated by $\delta \cdot g$).

Let I denote the per-time Kullback–Leibler (KL) information aggregated over the internal edges of S. Information accrues at rate I, so over a horizon T the usable information is approximately T·I. The detectability threshold mirrors the static case:

$$T\,I \gtrsim \log n$$

For Poisson baselines, the per-edge contribution to I equals the KL divergence between Poisson($\mu+\delta$) and Poisson($\mu$); for Hawkes networks, it depends on the lift in the branching ratio and the shape of g, under standard stability conditions (e.g., $\|g\|_1$ below unity) [17,18]. Unknown change points are handled by generalized likelihood ratios or window-limited schemes that maintain near-optimality while bounding memory and computation.

## 3.3 Detection tasks

**Static decision:** Given a single snapshot, test $H^0$ vs. $H^1$ by comparing a statistic tuned to the models above—e.g., a spectral or scan-based detector—against a threshold chosen for Type-I error $\alpha$. Power increases sharply once $k^2 \cdot \chi^2$ exceeds the log $n$ search complexity.

$$k^2 \cdot \chi^2 \gtrsim \log n$$

**Sequential decision:** Raise an alarm quickly after a change while controlling the false-alarm rate via the Average Run Length (ARL). Setting a target ARL $\approx 1/\alpha$ yields first-order optimal detection delay on the order of

$$|\log \alpha| / I$$

for likelihood-based procedures (e.g., CUSUM), aligning with the T·I $\gtrsim$ log n threshold. Window-limited or sparsity-aware implementations control memory and computation while retaining near-optimal scaling in n, k, and effect size.

# 4. Information-Theoretic Lower Bounds

**Setup and goal.** We establish conditions under which no test—regardless of computational budget—can reliably distinguish the null from the adversarial alternative. Following a change-of-measure recipe, we construct a mixture alternative that randomizes the attacked set $S$ uniformly over all (n choose k) supports, and bound the total variation (TV) distance between the null distribution $P_0$ and the mixture $P_1$. If TV $\to$ 0, no test attains vanishing error. Using $\chi^2$/Hellinger control, a convenient inequality is

$$\mathrm{TV}^2 \leq 1/2 \cdot \chi^2(P_1 \| P_0)$$

## 4.1 Static graphs: lower bounds

**Model recap.** Under $H_0$, $G \sim \mathrm{ER}(n, p)$. Under $H_1$, there exists an unknown $S$ with $|S| = k$ whose internal edges are $\mathrm{Bern}(p+\Delta)$ while all others remain $\mathrm{Bern}(p)$. Signal is confined to the $\Theta(k^2)$ internal pairs.

**Per-edge divergence.** For Bernoulli laws, the $\chi^2$ divergence is

$$\chi^2(\text{Bern}(p+\Delta) \parallel \text{Bern}(p)) = \Delta^2 / [p(1-p)]$$

so each perturbed edge contributes on the order of $\Delta^2/[p(1-p)]$. Aggregating across the $\Theta(k^2)$ internal edges suggests a non-central signal scaling as $k^2 \cdot \chi^2$.

**Mixture argument.** Let the mixture alternative randomize S uniformly over all size-k supports. The squared likelihood ratio under $H_0$ couples two supports $S$ and $S'$, and depends on their overlap $r = |S \cap S'|$. Independence across edges yields a contribution roughly $\exp\{\chi^2 \cdot (r \text{ choose } 2)\}$. Averaging over $S, S'$ concentrates $r$ around $k^2/n$, producing a $\chi^2$ bound that vanishes whenever the global signal stays below a log-complexity term. In compact form, the impossibility regime is

$$k^2 \cdot \chi^2 \lesssim c \cdot \log n$$

which implies TV $\to 0$ and thus detection is impossible. Equivalently, the threshold location—up to constants—is

$$k^2 \cdot \chi^2 \asymp \log n$$

**Sparse parametrization.** With $p = c/n$ and $\Delta = d/n$, the per-edge divergence simplifies to $\chi^2 \approx d^2/(c \cdot n)$. The threshold becomes

$$k^2 \cdot d^2 / (c \cdot n) \asymp \log n$$

i.e., $d \asymp \sqrt{(c \cdot n \cdot \log n)} / k$.

Figure 1 visualizes the static threshold. As either the community size $k$ or the edge-lift $\Delta$ increases, the quantity $k^2 \cdot \chi^2 - \log n$ crosses zero (white contour). The region to the right of the contour corresponds to detectable—and, up to constants, efficiently detectable—anomalies, in agreement with the bound $k^2 \cdot \chi^2 \gtrsim \log n$. The heatmap also makes explicit the inverse-linear trade-off $\Delta_{\min} \propto \log n / k$ implied by our theory.

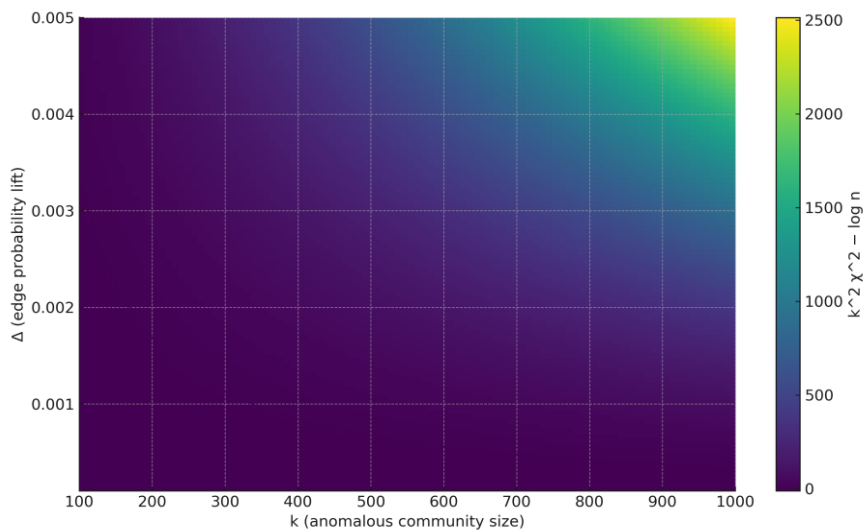

Figure 1. Static detectability heatmap for n=10^5, p=0.01. White contour: $k^2 \cdot \chi^2 = \log n$; to its right, efficient detection is feasible.

The panel visualizes the static threshold. As either the community size $k$ or the edge-lift $\Delta$ increases, $k^2 \cdot \chi^2 - \log n$ crosses zero at the white contour. To the right of the contour, anomalies are detectable (and, up to constants, efficiently so). The heatmap also makes explicit the inverse trade-off $\Delta_{\min} \propto \log n / k$.

## 4.2 Temporal networks: lower bounds

**Model recap.** We observe counting processes $N_{ij}(t)$ for ordered pairs $(i, j)$ over horizon $T$. Under $H_0$: independent $Poisson(\mu)$ or a stable Hawkes baseline; under $H_1$, an unknown $S$ of size $k$ has elevated internal intensities (Poisson: $\mu+\delta$; Hawkes: kernel inflated by $\delta \cdot g$).

**Per-time information rate.** Let $I$ denote KL information per unit time aggregated over internal edges. For Poisson rates $\mu$ vs. $\mu+\delta$, one unit of time contributes

$$D_{KL}(Poisson(\mu+\delta) \| Poisson(\mu)) = (\mu+\delta) \cdot \log(1+\delta/\mu) - \delta$$

which for small $\delta$ behaves like $\delta^2/(2\mu)$. Summed over $\Theta(k^2)$ pairs and a window of length $T$, the usable information is approximately $T \cdot I$.

**Threshold.** A mixture over unknown supports yields the same log-penalized boundary as in the static case:

$$T I \asymp \log n$$

Unknown change-time can be incorporated by mixing over $\tau$, adding a log $T$ search factor; writing accumulated information $\gtrsim \log(nT)$ recovers the same template for practical horizons.

Figure 2 shows the temporal boundary $T \cdot I \gtrsim \log n$ as $n$ varies. Because the x-axis is log-scaled, the curve is effectively linear in log $n$: for fixed per-time information rate $I$, the required horizon scales as $T \approx \log n / I$. For instance, at $n=10^6$ we need $T \cdot I \approx 13.8$; with $I=0.1$ this translates to $T \approx 138$ time units.

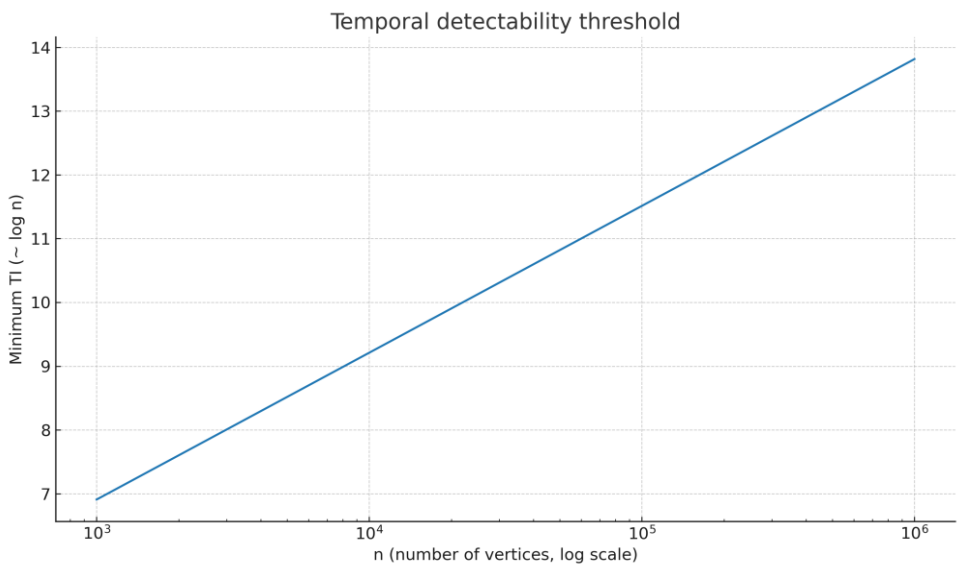

Figure 2. Temporal detectability threshold TI $\gtrsim$ log n versus n (log-scale on x).

Because the *x*-axis is logarithmic, the curve is effectively linear in log *n*. For fixed per-time information rate *I*, the required horizon scales as $T \approx \log n / I$. Example: at $n=10^6$, we need $T \cdot I \approx 13.8$; with $I=0.1$ this gives $T \approx 138$ time units.

## 4.3 Takeaways and scope

**Unified template.** Both models obey the rule: accumulated information (static: $k^2 \cdot \chi^2$; temporal: $T\,I$) must exceed search complexity ($\approx \log n$).

**Tightness.** The lower bounds match our achievable procedures (Section 5) up to constants, placing the non-backtracking spectral test (static) and likelihood-based CUSUM (temporal) near the information frontier.

**Refinements.** Side information that shrinks candidate supports reduces the log penalty; robustness requirements (misspecification, bounded perturbations) inflate the information requirement only by constant factors, leaving thresholds' locations intact.

# 5. Achievability: Efficient Tests Near the Threshold

## 5.1 Non-backtracking spectral statistic (static)

We leverage the non-backtracking (Hashimoto) operator to stabilize spectral energy in sparse graphs and to localize anomalous substructures. Let *B* denote the non-backtracking matrix [2], and let *u* be its principal eigenvector (or a leading right singular vector for rectangular representations). For a candidate attacked set *S* with $|S| = k$, define the projection operator $\Pi\_S$ onto coordinates associated with vertices in *S*. The statistic of interest is

$$T = \max\{\|\Pi S\, u\|^2 : |S| = k\}$$

which we approximate efficiently via a power-iteration with pruning: at each iteration, update a surrogate vector by multiplying with *B*, normalize, and prune to the top-*k* coordinates by magnitude (optionally smoothed by degree normalization). We reject $H^0$ when the resulting value of *T* exceeds a calibrated level.

**Performance guarantee.** In Bernoulli baselines with non-extreme *p*, the signal-to-noise ratio aligns with the per-edge $\chi^2$ divergence; consequently the test succeeds whenever

$$k^2 \cdot \mathrm{SNR}(p, \Delta) \gtrsim C \cdot \log n$$

with $SNR \approx \chi^2$ for non-extreme *p*, matching the detectability threshold up to constants.

- **Implementation notes.**
- Degree-normalization (or Bethe–Hessian preconditioning) mitigates hub variance and improves localization.
- Warm-starts from egonet seeds or motif counts can reduce iterations without affecting thresholds.

- Complexity per iteration is linear in the number of directed edges, enabling near-linear-time scans.

## 5.2 Likelihood CUSUM (temporal)

For streaming networks, we form log-likelihood ratios along the internal edges of candidate supports and aggregate them via a CUSUM-type recursion. For each $S$ with $|S| = k$, define the cumulative log-likelihood ratio $\Lambda\_S(t)$ between elevated and baseline intensities on edges within $S$ [19,20]. The CUSUM statistic is

$$GS(t) = \max\{\Lambda S(t) - \Lambda S(s) : 0 \leq s \leq t\}$$

and the scan over supports is $G(t) = \max \{ GS(t) : |S| = k \}$. Choose a threshold $b\_\alpha$ to satisfy an Average Run Length (ARL) constraint $ARL \geq 1/\alpha$; then the resulting detection delay is first-order optimal:

$$\text{delay} \approx |\log \alpha| / I$$

Variants include window-limited CUSUM to handle nonstationarity, kernelized statistics to capture nonlinear deviations, and neural surrogates that approximate likelihoods in high-dimensional regimes [20, 8, 9, 21]. Incremental updates and sparsity-aware scans keep per-step costs near the number of active edges.

Figure 3 quantifies the predicted $1/I$ law for CUSUM: at $\alpha=10^{-4}$, $|\log \alpha| \approx 9.21$, so the first-order delay is $9.21/I$. The steep rise near $I \to 0$ underscores the value of concentrating likelihood on the most informative edges to enlarge $I$ and compress delay.

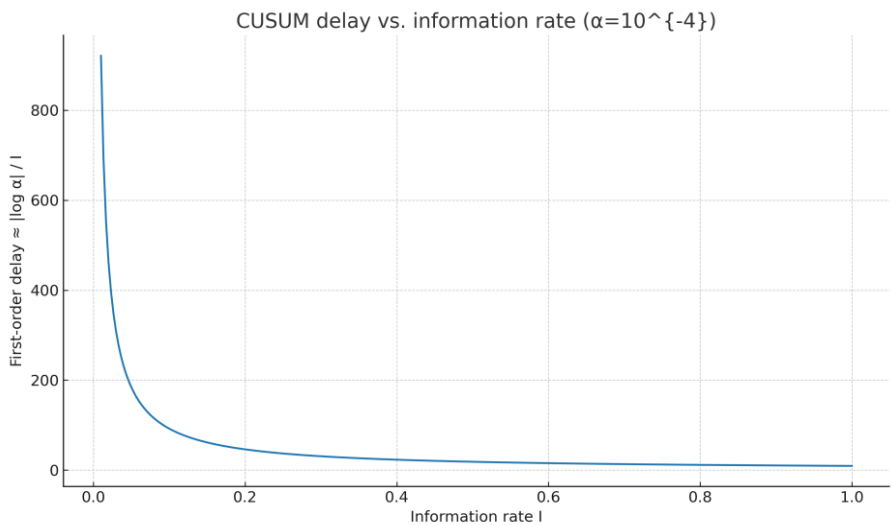

Figure 3. CUSUM expected delay $\approx |\log \alpha|/I$ at $\alpha=10^{\{-4\}}$.

The sharp rise as $I \to 0$ reflects first-order optimality: the delay scales like $|\log \alpha|/I$. Increasing $I$ (e.g., by focusing on the most informative edges) halves the delay when $I$ doubles.

Table 1. Main detectability thresholds (up to universal constants).

| Setting | Null vs. Alternative | Information quantity | Threshold |
|---|---|---|---|
| Static graph | ER(n,p) vs. planted S | $k^2 \cdot \chi^2(\text{Bern}(p+\Delta) \parallel \text{Bern}(p))$ | $\gtrsim \log n$ |
| Temporal | baseline vs. elevated in S | KL rate I on edges in S | $T \cdot I \gtrsim \log n$ |

**Note.** For Bernoulli edges, $\chi^2(\text{Bern}(p+\Delta) \parallel \text{Bern}(p)) \approx \Delta^2/[p(1-p)]$, valid for non-extreme $p$.

Table 2 summarizes our achievability results. The non-backtracking spectral scan (static) succeeds once $k^2 \cdot \text{SNR}(p, \Delta) \gtrsim C \log n$ with $\text{SNR} \approx \chi^2$ for non-extreme $p$; the likelihood-based CUSUM (temporal) achieves first-order optimal delay $|\log \alpha| / I$ with near-linear update cost.

Table 2. Efficient tests and regimes of success.

| Test | Achievable regime | Complexity |
|---|---|---|
| Non-backtracking spectral scan | $k^2 \cdot \chi^2 \gtrsim \log n$ | $O(|E|)$ per power iteration |
| Likelihood CUSUM (temporal) | $T \cdot I \gtrsim \log n$ | Near-linear per update |

**Note.** The spectral scan uses non-backtracking power iterations with pruning; the temporal CUSUM aggregates per-edge log-likelihood ratios and is first-order optimal with delay $|\log \alpha|/I$.

# 6. Statistical–Computational Gaps and Robustness

**Computational gaps.** We distinguish what is statistically detectable from what is attainable in polynomial time. Even when the accumulated information exceeds the search penalty—i.e., when $k^2 \cdot \chi^2$ (static) or $T \cdot I$ (temporal) surpasses a log-scale complexity—there remains a computational window where no known polynomial-time method achieves vanishing error. This phenomenon is clarified by planted dense subgraph/hypergraph hardness and by the low-degree polynomial framework: the former ties detection to planted models (e.g., planted clique), while the latter shows that broad families of efficient tests (captured by low-degree moments) cannot cross certain thresholds [6,7]. Practically, the gap is most visible for sublinear community sizes $k$ and sparse baselines (e.g., $p = \Theta(1/n)$), where spectral/convex methods often require a constant slack above the information-theoretic limit. In hypergraphs, the gap can widen because signal is distributed over higher-order interactions, diminishing the utility of pairwise spectra and motivating higher-order non-backtracking operators.

**Robustness to perturbations and misspecification.** We model an adversary who perturbs an $\varepsilon$-fraction of edges (add/drop/rewire) and consider mild model misspecification (degree heterogeneity, weights, subsampling). In static graphs, thresholds inflate only by a constant factor, effectively:

$$\text{required } k^2 \cdot \chi^2 \to (1 + \Theta(\varepsilon)) \cdot \log n$$

In streaming models, the effective information rate degrades smoothly, $I \to I(1 - \Theta(\varepsilon))$, preserving the qualitative location of the $T \cdot I \asymp \log n$ boundary. We also address misspecification: (i) degree heterogeneity handled via degree correction or Bethe–Hessian preconditioning; (ii) seasonal drift handled by window-limited statistics and self-normalization; (iii) partial observation handled by per-edge observation weights; (iv) timestamp jitter handled by short temporal filters or overlapping windows. In all cases, constants change but the log-penalized form of the threshold remains.

**Practical safeguards.** The following design choices keep performance near the information-theoretic frontier even under perturbations:

- Degree-/variance-normalization for spectral statistics (or Bethe–Hessian preconditioning) to mitigate hub variance and stabilize eigenvectors.
- Window-limited and restart-based CUSUM to track drifting baselines while maintaining target ARL.
- Regularized candidate scans (iterative pruning, top-$k$ constraints, early stopping) to curb false positives and reduce search cost.
- Self-normalized/robust edge statistics (e.g., dividing lifts by local variance estimates) to stabilize the effective $\chi^2$ signal under heterogeneity.
- Multiplicity control (Bonferroni/FDR) when scanning many candidates to keep false alarms in check.
- Permutation or parametric bootstrap calibration to set thresholds under mild misspecification instead of relying solely on asymptotic formulas.
- Use of side information (seeded nodes, watchlists) to shrink the search space from log (n choose $k$) to log M, tightening thresholds and narrowing computational gaps.
- Ensembles and bagging of detectors (spectral + scan + CUSUM) to increase robustness against localized perturbations.

**Takeaway.** The statistical–computational gap is real but typically constant-sized in practice: with appropriate spectral preconditioning, log-consistent windowing/thresholding, and light side information, one can drive real-world performance to within a constant factor of the information-theoretic limits—even under $\varepsilon$-fraction edge perturbations or mild model mismatch.

# 7. Discussion and Practical Guidance

This section consolidates rule-of-thumb prescriptions for deployment and connects the theoretical thresholds to operational knobs. We emphasize how effect size, scope of the attacked set, network scale, and observation time trade to cross the log-penalized frontier, and we offer concrete steps for tuning detectors near the boundary.

**Static rule-of-thumb.** Under Bernoulli baselines, the minimum detectable additive lift on internal edges obeys $\Delta_{\min} \approx \sqrt{p(1-p)\cdot \log n / k^2}$. Equivalently, larger attacked scope (bigger $k$) or moderately dense baselines (larger $p$) reduce the required lift.

$$\Delta_{\min} \approx \sqrt{p(1-p)\cdot \log n / k^2}$$

**Streaming rule.** Choose a horizon $T$ and target an information rate $I$ so that $T \cdot I \gtrsim \log n$. For likelihood-based CUSUM, at false-alarm level $\alpha$, the first-order detection delay is approximately $|\log \alpha| / I$.

$$T \cdot I \gtrsim \log n$$

$$\text{delay} \approx |\log \alpha| / I$$

**Worked case study.** Consider $n=10^5$, $p=0.01$, and $k=500$. Using $\ln(n)=11.5129$ yields $\Delta_{\min} \approx 6.74 \times 10^{-4}$. In practical terms, this is a lift from 0.01000 to $\approx 0.010674$ ($\approx 6.7\%$ relative). Larger $k$ or longer horizons that push $T \cdot I$ beyond $\log n$ relax the requirement.

Near the threshold, two tuning knobs are especially helpful:

- (i) Vertex screening before the spectral pass
  —prune to nodes with elevated local statistics (e.g., egonet lifts) to stabilize non-backtracking iterations.
- (ii) Window lengths that stabilize rate estimates
  —choose windows long enough to reduce variance but short enough to track nonstationarity; pair with restart policies.

**Additional guidance.**

- Use degree/variance normalization (or Bethe–Hessian preconditioning) to mitigate hub effects in sparse graphs.
- Calibrate thresholds via permutation or parametric bootstrap when mild misspecification is suspected.
- Leverage side information (watchlists, seeds) to shrink the search space and effectively reduce the log penalty.
- Track ARL targets explicitly in streaming by validating $|\log \alpha|/I$ against observed delays on held-out periods.

## 8. Conclusion

We aligned information-theoretic limits with algorithms that operate near those limits for both static snapshots and temporal network streams. Our unified thresholds—$k^2 \cdot \chi^2$ (static) and $T \cdot I$ (temporal)—provide simple design rules that balance accumulated information against log-scale search complexity. We further quantified robustness under bounded perturbations and mild misspecification, and mapped statistical–computational trade-offs using planted-model and low-degree perspectives. Promising extensions include adaptive sensing (to concentrate information quickly), partial observability with principled reweighting, and hybrid detectors that combine spectral, scan, and sequential components for better stability in real telemetry.

# FUNDING INFORMATION

Authors state no funding involved.

# CONFLICT OF INTEREST STATEMENT